\documentclass{sig-alternate-2013}

\newfont{\mycrnotice}{ptmr8t at 7pt}
\newfont{\myconfname}{ptmri8t at 7pt}
\let\crnotice\mycrnotice%
\let\confname\myconfname%

\toappear{CompSci'13, October 28, 2013, San Francisco, CA, USA.}

\usepackage{times}
\usepackage{hyperref}
\usepackage{algorithmic}
\usepackage{amsmath}
\usepackage[tight,footnotesize]{subfigure}
\usepackage{graphicx}
\usepackage{color}
\usepackage{array}
\usepackage{colortbl}
\usepackage{url}

\title{Revealing Comparative Advantages \\ in the Backbone of Science}

\numberofauthors{2}
\author{
\alignauthor Miguel Guevara\\
	   \affaddr{Department of Informatics}\\
       \affaddr{Universidad de Playa Ancha}\\
       \affaddr{Valpara\'iso, Chile}\\
       \email{miguel.guevara@upla.cl} \\
\alignauthor Marcelo Mendoza\\
	   \affaddr{Department of Informatics}\\
       \affaddr{Universidad T\'ecnica Federico}\\
       \affaddr{Santa Mar\'ia}\\
       \affaddr{Santiago, Chile}\\
       \email{marcelo.mendoza@usm.cl} \\
}

\begin{document}
\maketitle


\begin{abstract}
Mapping Science across countries is a challenging task in the field of Scientometrics.
A number of efforts trying to cope with this task has been discussed in the state of the art, 
addressing this challenge by processing collections of scientific digital libraries and 
visualizing author-based measures (for instance, the h-index) or document-based measures 
(for instance, the averaged number of citations per document). A major drawback of these approaches 
is related to the presence of bias. The bigger the country, the higher the measure value. 
We explore the use of an econometric index to tackle this limitation, known as the Revealed 
Comparative Advantage measure (RCA). Using RCA, the diversity and ubiquity of each field of knowledge 
is mapped across countries. Then, a RCA-based proximity function is explored to visualize citation and h-index ubiquity. 
Science maps relating 27 knowledge areas and 237 countries are introduced using data crawled from Scimago that 
ranges from 1996 to 2011. Our results shows that the proposal is feasible and can be extended to 
ellaborate a global scientific production characterization.
\end{abstract}




\section{Introduction}
Mapping Science is a task that can be backtracked to the seminal work of Moreno in 
1934 \cite{moreno34} where the first sociogram was introduced. Later, the 
first \emph{Map of Science} was proposed by Bernal in 1939 \cite{bernal39}. Since then, 
a number of efforts supported by the theoretical framework of Network Science has been discussed, 
introducing maps that try to reveal the complexity of Science.

Scientometrics, or the Science of Science analysis, cope with scientific impact measures and 
their relationships to Science success. On the one hand, docuemtn impact measures try to model how 
the content of a specific article influences a scientific community. Measures as the number of 
citations or the impact factor of the journal that publish the article~\cite{garfield55} are widely 
used to evaluate this dimension. On the other hand, authors impact measures try to quantify the 
influence of a given author in a scientific community. Measures as the total amount of citations 
per author or the h-index~\cite{hirsch05} are widely used for this purpose.

Science maps are visual representations of scientific artifacts and their relationships.
Previous efforts are focused on the visualization of the structure of Science, looking for maps 
that unleash relationships between disciplines and/or authors. In this direction, 
Katy B\"{o}rner~\cite{borner_atlas_2010} compiled a number of Science maps, providing a 
coherent corpus named ``Atlas of Science''. The Atlas shows that several efforts are based on 
the use of proximity functions applied to disciplines, where the proximity is related to 
citations between documents or to collaborations between authors~\cite{boyack_mapping_2005}.
Among these efforts, there are some of them that map each scientific artifact to countries, 
creating geolocated visualizations of science objects. In this direction, the paper 
authored by Batty~\cite{batty03} explore the geography of citations in the ISI database, 
visualizing the amount of citations across cities in USA, Europe and Japan. 
These maps exhibit evidence of bias. Citations are biased to economic complexity, 
illustrating that evolved cities and scientific production correlates. Bias is a main drawback 
for mapping science across countries. Whilst big countries may be overrepresented in this kind 
of visualizations, small countries may dissapear. To tackle this problem, we explore the use 
of an econometric index, known as the Revealed Comparative Advantage (RCA)~\cite{rca}. 
The RCA measure calculates the relative advantage/disadvantage of a country for a 
specific product in correspondence to its own internal market.
In this direction, Hidalgo~\cite{hidalgo_product_2007} showed that the RCA measure 
exhibits good properties for the characterization of the economic complexity of countries, 
favoring fair comparisons between countries of different sizes.

In this article we explore the use of RCA for the characterization of scientific production, 
mapping science across countries. Using RCA, we include the diversity and ubiquity of each knowledge 
field for each country. Then, a RCA-based proximity function is explored to visualize citation 
and h-index ubiquity networks. Science maps relating 27 knowledge areas and 237 countries are 
introduced using data crawled from Scopus that ranges from 1996 to 2011. Experimental results show 
that the proposed approach is feasible.

The remainder of the paper is organized as follows. 
A summary of the related work is discussed in Section 2.
In Section 3 we introduce the RCA framework. 
Experimental results are shown in Section 4.  
Finally we conclude in Section 5.

\section{Related work}

Geography of Science and Scientific Wealth-Impact of Nations \cite[p. 203]{borner_atlas_2010}, 
are very close topics to our work, in the sense that our approach attempts to relate both. 
On one hand, Geography of Science is linked with works aimed to understand the geographic 
distribution of citations and relationships between countries or institutions in development of Science. 
In this direction, the work of Batty~\cite{batty03} revels a concentration pattern over scientist, 
institutions and, in consequence countries. He shows that only 27 countries concentrates 
the world scientific production. In the same sense, Carvalho and Batty~\cite{carvalho_geography_2006} 
explore the geography of scientific production in USA in the area of Computer Science, 
finding that the productivity of USA research centers is highly skewed. On the other hand, 
Scientific Wealth-Impact of Nations~\cite{may_scientific_1997, king_scientific_2004} is linked to works 
that analyze the development and capabilities of countries or regions in the scientific map. 
May \cite{may_scientific_1997} presents a scientific research outputs among several developed 
countries discussing the impact of Asian nations in the global science production. 
Zhou and Gl\"an~\cite{zhou-depth2010} present an analysis of the contribution and collaboration 
of China with other nations to produce scientific documents. As our work, they also use an RCA-based index 
but only related to the production of papers to measure the impact on country activity. 
A recent technical report of The Royal Society~\cite{the_royal_society_knowledge_2011} presents 
an extensive analysis about emerging scientific nations, incomes and policies around the world. 
They also present collaboration between nations using cite-oriented networks. They also try to explain 
which country is leading the scientific development and who is at the bottom of the ranking list. 

The main difference between the related work and our proposal is that we provide a framework that makes 
possible the analysis of each country in the global context, comparing countries of different sizes and complexities, 
favoring the creation of a global scientific production characterization. 

\section{The Science Production Model} \label{sec:rca}


To create a Science production model we take into account the relative size between small and big countries, 
tackling the bias introduced in the analysis by country size. We do this by modeling the scientific production of a 
country as a relative measure of its internal diversity and the global ubiquity of the discipline where 
the scientific product is generated. We do this by using an econometric index known as the Revealed Comparative 
Advantage (RCA) to model country diversity and product ubiquity. RCA is an index used in economics that allows 
to calculate relative advantages of a country for a specific product.  

A comparative advantage refers to the ability of a country to produce a specific product at a 
lower marginal and opportunity cost over another. As different countries have different relative efficiencies, 
every country achieves a gain by collaborating with each other. We take advantage of this key factor 
for the analysis of scientific production at a country level.

We measure comparative advantages by using the RCA measure. 
RCA raises that a country has a comparative advantage in a product if it produces more 
than the total world trade that the product represents. 
We model scientific production by analogy with world trade. 
For example, Computer Science produces 1,222,228 citable documents in Scimago from 1996 to 2011, 
representing 4\% of world scientific production. Of this total, Hong Kong produced 16,684 
documents~\footnote{A document belongs to a country if at least one author is affiliated to an institution of this country.}, 
and since Hong Kong's total scientific production contributes with 174,400 documents, Computer Science accounted for 
9.5\% of Hong Kong's scientific production. This represents more than 2 times Computer Science's world trade, 
so we can say that Hong Kong has a comparative advantage in Computer Science.

Notice that RCA is not constrained to the amount of citable documents. 
We can replace citable documents by another scientific production index, repeating the analysis. 
Thus, RCA introduces a methodological framework for the revealing of comparative advantages over a 
number of scientific production indexes, allowing comparisons based on, for instance, 
number of documents, citations or h-index. Accordingly, we will formalize the RCA-based analysis 
for a generic index of scientific production.  

Let $X_{c,f}$ be a generic index that represents the production of a country $c$ in a specific field of knowledge $f$. 
The RCA that a country $c$ has for the field of knowledge $f$ is given by the following expression:

\[
\tt{RCA}_{c,f}=\frac{X_{c,f}/\sum_{f}X_{c,f}}{\sum_{c}X_{c,f}/\sum_{c}\sum_{p}X_{c,f}}, 
\]

\noindent which measures whether a country $c$ produces more in a field of knowledge $f$, as a 
share of its total scientific production ($\sum_{f}X_{c,f}$), than the average global production 
for $f (\sum_{c}X_{c,f}/\sum_{c}\sum_{p}X_{c,f})$.

We can characterize the global scientific production by calculating 
$\tt{RCA}_{c,f}$ for every country $\times$ field of knowledge pair. We create a binary matrix 
that represents comparative advantages in the country $\times$ field of knowledge space by setting 
entries in the matrix to 1 for comparative advantages pairs. Formally, let $M_{c,f}$ an entry of the 
matrix of comparative advantages. $M_{c,f}$ is defined by the following equation:

\[
M_{c,f}= \left\{
\begin{array}{ll}
         1 & \mbox{if $\tt{RCA}_{c,f} \geq 1$} , \\
         0 & \mbox{otherwise}
\end{array}
\right .
\]

Thus, the matrix of $M_{c,f}$ entries represents a knowledge space characterized from revealed 
comparative advantages between every country $\times$ field of knowledge pair. The knowledge space 
summarizes which country produces what and who is contributing to every field of knowledge. 
Accordingly, we can measure diversity (from a country point of view) and ubiquity (from a field of 
knowledge point of view) by summing rows or columns of the knowledge space matrix, respectively.

Formally, let $\mbox{Div}_{c}$ be a diversity measure for a certain country $c$, and 
let $\mbox{Ubi}_{f}$ be an ubiquity measure for a certain field of knowledge $f$. We calculate 
$\mbox{Div}_{c}$ and $\mbox{Ubi}_{f}$ by using $M_{c,f}$ entries as follows:

\[
\tt{Div}_{c} = \sum_f M_{c,f}, \hspace{5mm} \tt{Ubi}_{f} = \sum_c M_{c,f}.
\]

Now we can introduce the concept of proximity between fields of knowledge. 
A proximity notion between different fields $f_1$ and $f_2$ can be calculated as 
the minimum of the pairwise conditional probability of producing in a field of knowledge 
given that it produces in the other:

\[
\phi_{f_1,f_2}(c) = \tt{Min} \{ P(\tt{RCA}_{c,f_1} \mid \tt{RCA}_{c,f_2}) , 
P(\tt{RCA}_{c,f_2} \mid \tt{RCA}_{c,f_1}) \}.
\]
 
Notice that RCA conditionals define an asymmetric relation between fields of knowledge. 
Then, the use of the minimum introduces symmetry, taking the minimum conditional value of the pair. 
The use of the minimum reduces the false positive rate in the knowledge space. 

Now we can estimate these probabilities by taking into account the entries of the knowledge space matrix: 

\[
\phi_{f_1,f_2} = \frac{\sum_c M_{c,f_1} \cdot M_{c,f_2}}{\tt{Max} \{ \tt{Ubi}_{f_1} , \tt{Ubi}_{f_2} \}  }.
\]

We illustrate how this proximity function works by using an example. 
Scimago shows that during the period ranged from 1996 to 2011, 37 countries 
produced knowledge in Computer Science, because they achieved a production in number of citable 
documents greater than the world trade in this field of knowledge (4\%). Analogously, 
54 countries produced knowledge in Decision Sciences, with a production greater than 
the 0.4\% world trade in this field, and 28 countries produced both. 
Then, the proximity between both fields is $\tfrac{28}{54} = 0.51$. 
Notice that the Decision Sciences ubiquity is greater than the Computer Science one, 
reason why we divide by 54 instead of 37. 

Note that we can apply the proximity function to compare pairs of countries. 
This can be done by aggregating the knowledge space matrix entries across fields of knowledge 
and by using diversities instead of ubiquities, as follows:

\[
\phi_{c_1,c_2} = \frac{\sum_f M_{c_1,f} \cdot M_{c_2,f}}{\tt{Max} \{ \tt{Div}_{c_1} , \tt{Div}_{c_2} \}  }.
\]

Regarding number of citable documents, Scimago shows that during the period ranged from 1996 to 2011, 
USA produced in 15 fields of knowledge, China produced in 11 and both countries had 2 fields of knowledge 
in common (Business, Management and Account, and Multidisciplinary). Then, the proximity between both 
countries is $\tfrac{2}{15} = 0.13$. Note that as USA is more diverse than China in the production of science, 
producing in 15 fields of knowledge over a total of 27 fields of knowledge considered in the dataset, 
we divide by 15 instead of 11. 



\vspace{5mm}

\section{Results}
\subsection{Dataset and Nomenclature}
We crawled the data from the SCImago Journal \& Country Rank site~\cite{scimago_scimago_2007}. 
27 fields of knowledge were crawled summarizing the scientific production of 237 countries in the period 
that ranges from 1996 to 2011. For each field of knowledge $\times$ country pair, 
the following indexes were retrieved: citable documents, citations, self citations, 
citations per document (in average), and H-index. We interpret these indexes by analogy with 
consumer-producer relationships. Document citations can be seen as an analogy for consumption. 
Document creation can be seen as an analogy for production. 
A summary of some statistics of the dataset is shown in Table~\ref{tab:resume}.

\begin{table}[htdp]

\begin{center}
\begin{tabular}{|l|c|} \hline
	Period & 1996-2011 \\ \hline
	Countries & 238 \\ \hline
	Fields of knowledge & 27 \\ \hline
	Citable documents & 29,895,499  \\ \hline 
	Citations & 429,922,232 \\ \hline  
	Self citations & 137,399,721 \\ \hline  
	$\sum$ citations per document & 55,578 \\ \hline
    $\sum$ H-index & 169,717 \\ \hline
\end{tabular}
\end{center}
\caption{The SCImago Journal and Country Rank dataset}
\label{tab:resume}
\end{table}

We use an abbreviation for each field of knowledge, that consists of a preffix. 
Table~\ref{tab:fields} denotes these abbreviations, that will be consistently used in our visualizations. 

\begin{table}[h!]
\small
\begin{center}
\begin{tabular}{|r|l|} \hline 
Field of Knowledge&Label \\ \hline \hline
Mathematics&Mth \\
Physics and Astronomy&Phy-Ast \\
Chemistry&Chm \\
Chemical Engineering&ChmEng \\
Multidisciplinary&Mlt \\
Agricultural and Biological Sciences&Agr-BlgScn \\
Earth and Planetary Sciences&Ert-PlnScn \\
Veterinary&Vtr \\
Energy&Enr \\
Environmental Science&EnvScn \\
Materials Science&MtrScn \\
Engineering&Eng \\
Economics, Econometrics and Finance&Ecn-Ecnm-Fnn \\
Business, Management and Accounting&Bsn-Mng-Acc \\
Social Sciences&SclScn \\
Arts and Humanities&Art-Hmn \\
Psychology&Psy \\
Decision Sciences&DcsSci \\
Computer Science&CmpScn \\
Neuroscience&Nrsc \\
Biochemistry, Genetics and Molecular Biology&Bch-Gnt-MlcBlg \\
Health Professions&HltPrf \\
Immunology and Microbiology&Inm-Mcr \\
Pharmacology, Toxicology and Pharmaceutics&Phr-Txc-Phr \\
Nursing&Nrs \\
Dentistry&Dnt \\
Medicine&Mdc \\ \hline

\end{tabular}
\end{center}
\caption{Fields of knowledge and their corresponding labels.}
\label{tab:fields}
\end{table}


\subsection{RCA Analysis}

We start this section by calculating RCA matrices.  
According to Section~\ref{sec:rca}, we calculate each matrix entry by replacing $X_{c,f}$ by 
number of documents, citations or H-index, obtaining three RCA-based matrices. 
The expected RCA value is 1, that is to say, it is expected that the average country produces 
in a certain field exactly the world trade. Thus, RCA values distribute around 1. 
In Table~\ref{tab:rca} we summarize some statistics for these distributions.

\begin{table}[htdp]
\small
\begin{center}
\begin{tabular}{|l|c|c|c|c|c|c|} \hline
RCA            & Min.  & 1st Qu. &   Median  &   Mean    &  3rd Qu. &   Max.  \\ \hline \hline
Documents      & 0.0   & 0.416   &   0.827   &   1.289   &  1.427   &   144.3 \\
Citations      & 0.0   & 0.271   &   0.741   &   1.349   &  1.420   &   110.0 \\
H-index        & 0.0   & 0.640   &   0.952   &   1.116   &  1.290   &    23.7 \\ \hline
\end{tabular}
\end{center}
\caption{RCA distributional statistics for number of documents, citations and H-index.}
\label{tab:rca}
\end{table}

The first and third quantiles suggest the symmetry or skewness of each distribution. 
Documents and citations RCA distributions are skewed because the third quartiles are farther 
above the median than the first quartiles. Note that this behavior can be also observed 
in the maximum values of both distributions. On the other hand, the H-index RCA distribution 
exhibits symmetry because the first and third quartiles are equally distant from the median. 
The maximum behaves the same way. 

Finally we explore the strength of the linear relationships between these distributions by measuring the 
Pearson correlation coefficient. The correlation for citations and number of documents is $r = 0.539$, 
for citations and H-index is $r = 0.681$ and for number of documents and H-index is $r = 0.632$, 
showing that the strongest linear relationship is between citations and H-index. 

\subsection{Diversity and Ubiquity}

Now we continue the analysis by exploring diversity and ubiquity indexes. 
Fields of knowledge can be characterized by using ubiquity measures, trying to 
understand how a certain field of knowledge is developed in different countries around the world. 
We calculate ubiquity according to three indexes: citable documents, citations and H-index. Ubiquity ranges 
from 0 to 237 indicating the number of countries that produces in the field of knowledge over the world 
trade of this field (RCA > 1). These results are depicted in Table~\ref{tab:ubi}.

\begin{table}[h!]
  \small
\begin{center}
\begin{tabular}{|l|c|c|c|} \hline 
Label	&	$\mbox{Ubi}$ Docs	&	$\mbox{Ubi}$ Citations	&	$\mbox{Ubi}$ H-index	\\
\hline \hline
\rowcolor[gray]{.85}
Agr-BlgScn	&	178	&	179	&	165	\\
Art-Hmn	&	76	&	54	&	79	\\
Bch-Gnt-MlcBlg	&	24	&	18	&	87	\\
Bsn-Mng-Acc	&	67	&	45	&	68	\\
ChmEng	&	53	&	68	&	74	\\
Chm	&	55	&	60	&	81	\\
CmpScn	&	37	&	44	&	64	\\
DcsSci	&	54	&	57	&	55	\\
Dnt	&	67	&	64	&	65	\\
\rowcolor[gray]{.85}
Ert-PlnScn	&	124	&	122	&	128	\\
Ecn-Ecnm-Fnn	&	82	&	52	&	79	\\
Enr	&	84	&	98	&	96	\\
Eng	&	38	&	58	&	83	\\
\rowcolor[gray]{.85}
EnvScn	&	172	&	166	&	153	\\
HltPrf	&	53	&	32	&	63	\\
\rowcolor[gray]{.85}
Inm-Mcr	&	141	&	130	&	120	\\
MtrScn	&	42	&	60	&	72	\\
Mth	&	72	&	79	&	80	\\
\rowcolor[gray]{.85}
Mdc	&	124	&	102	&	141	\\
Mlt	&	84	&	34	&	47	\\
Nrsc	&	34	&	24	&	59	\\
Nrs	&	75	&	46	&	69	\\
Phr-Txc-Phr	&	70	&	72	&	97	\\
Phy-Ast	&	53	&	59	&	71	\\
Psy	&	40	&	27	&	59	\\
\rowcolor[gray]{.85}
SclScn	&	126	&	115	&	139	\\
\rowcolor[gray]{.85}
Vtr	&	126	&	130	&	113	\\ \hline
\end{tabular}
\end{center}
\caption{Ubiquity indexes for each field of knowledge. Gray rows indicates the most ubiquity fields.}
\label{tab:ubi}
\end{table}

As Table~\ref{tab:ubi} shows, the most ubiquity fields of knowledge are 
``Agricultural and Biological Sciences'' and ``Environmental Science''. 
Note that there is a correspondence between number of citable documents, citations and H-index 
regading the ubiquity index, that is to say, more documents tend to correlate with more citations 
and more citations tend to correlate with high H-index values. On the other hand, the lowest 
ubiquity is showed by ``Biochemistry, Genetics and Molecular Biology''. Note that in this last case, 
the ubiquity of the H-index is 87, whilst their docs and citations ubiquities are 24 and 18, respectively, 
showing that the correspondence between ubiquities for low ubiquity fields is weak. Note that something 
similar occurs in ``Computer Science'', ``Engineering'' and ``Neuroscience'', with low ubiquity measures 
in citable documents and citations, but with significant values regarding the H-index. 
This fact can be explained by considering that in these fields only a few countries are competitive 
in terms of volume (number of documents or citations), but there are more countries that are competitive 
in terms of high cited papers. 

\subsection{Proximity Analysis} 

\begin{figure}[h!]
\centering
\includegraphics[width=6.5cm]{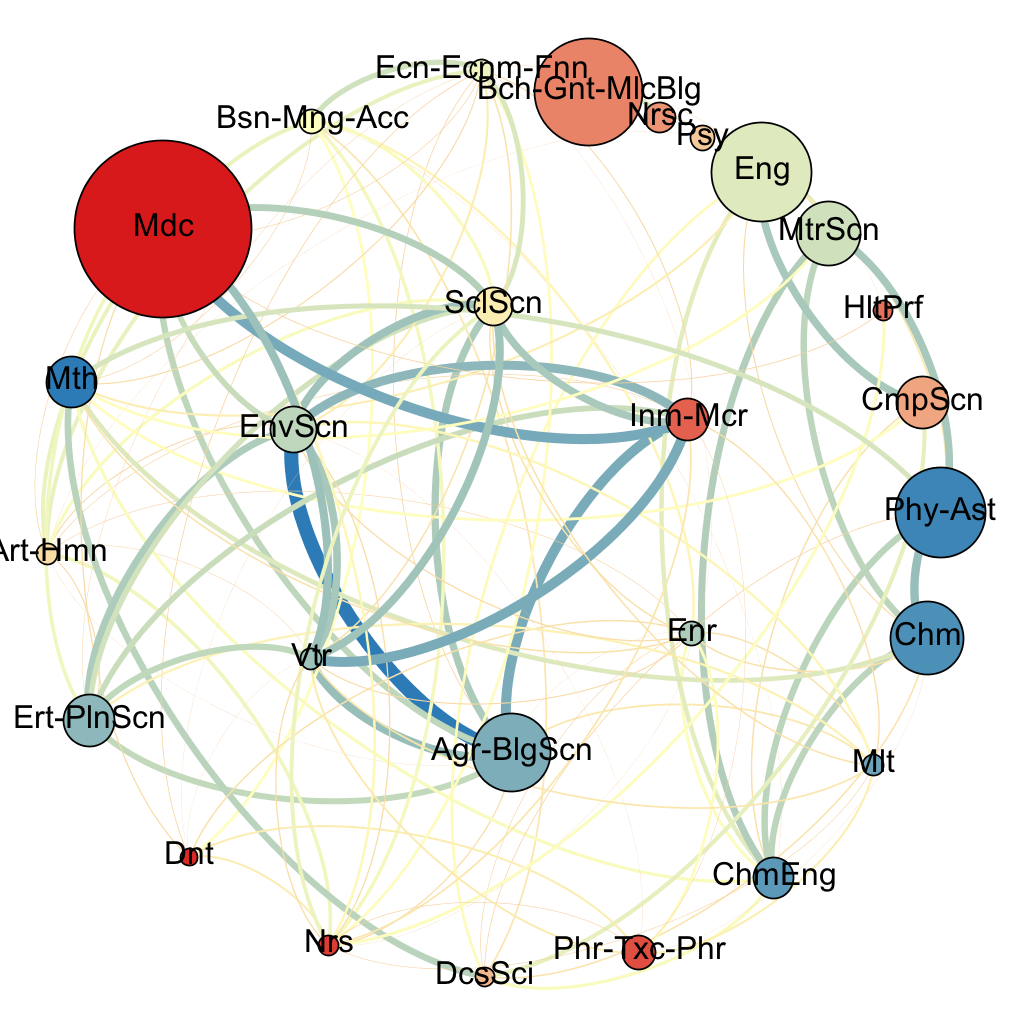}
\caption{A network visualization for RCA proximity based on number of citable documents. The counterclockwise order of the nodes is defined by the sum of the weights of each incoming edge but the size of the node is proportional to the number of citable documents of the field.}
\label{fig:citdocs}
\end{figure}

Now we explore RCA-based proximity for each scientific production index considered in this analysis. 
We start by measuring proximities between fields of knowledge pairs according to the number of citable documents. 
We show this network in Figure~\ref{fig:citdocs}.

We used a double circular layout, that has to be read in an counterclockwise, beginning at Maths (inner circle). 
For the remain of the paper, we maintain this color coding fashion and the same layout to visualize the 
knowledge space relationships.
As Figure~\ref{fig:citdocs} shows, the node order is related to the ubiquity but it is not related to 
the number of citable documents. This fact indicates that ubiquity and production volume are not related. 
In particular, we can observe that the field with more documents (``Medicine'') achieves the 9th place in 
terms of ubiquity. Moreover, ``Social Sciences'' and ``Environmental Science'' register few citable 
documents despite the fact that these fields are the most ubiquous. Note also that the closest 
fields of knowledge are ``Environmental Science'' and ``Agricultural and Biological Sciences'', 
suggesting a semantic relationship. Something similar is observed between ``Medicine'' and ``Inmunology''. 

\begin{figure}[h!]
\centering
\includegraphics[width=6.5cm]{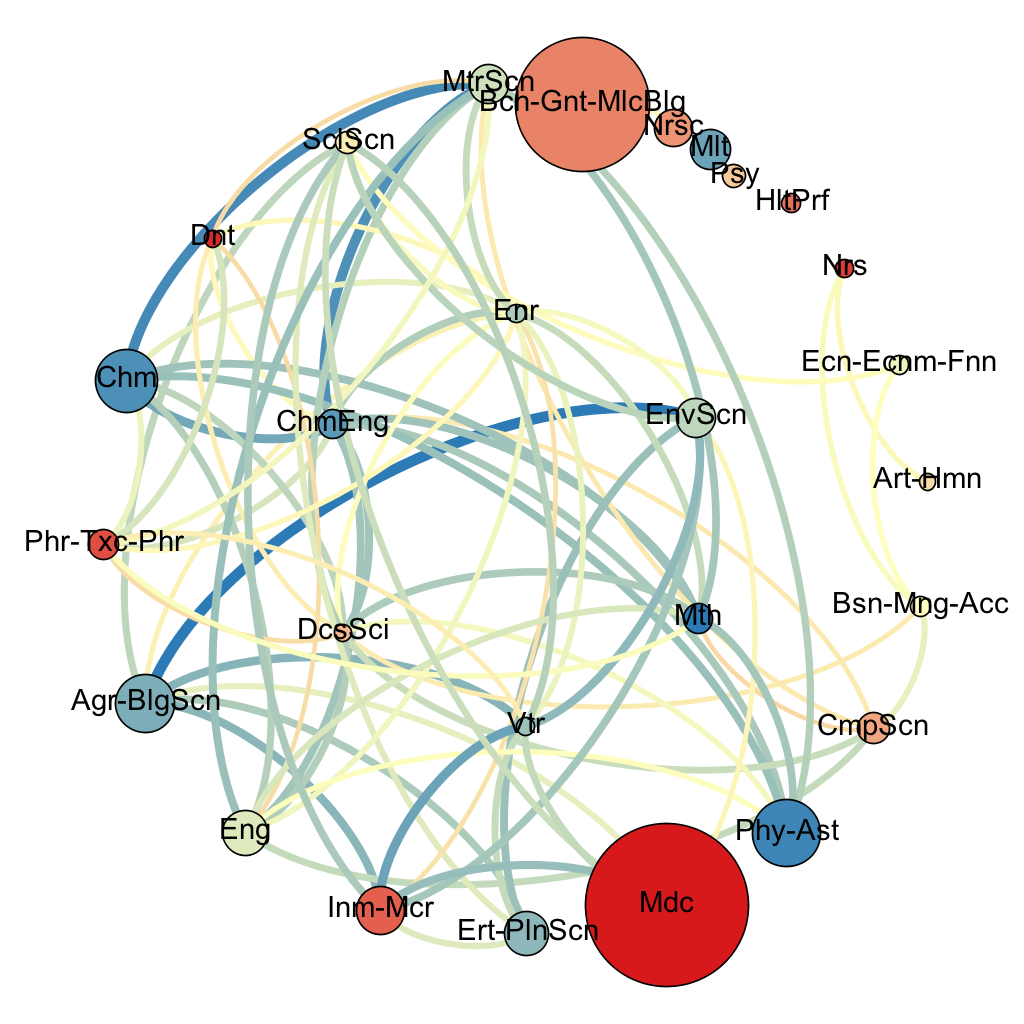}
\caption{A network visualization for RCA proximity based on number of citations.}
\label{fig:citations}
\end{figure}

We conduct a similar analysis by visualizing proximities between fields of knowledge based on number of citations. 
As in the previous network, the node order is defined by the sum of the weights of the incoming edges (that is to say, 
to the ubiquity of this index), and the node size is proportional to the number of citations of the field. 
This network is shown in Figure~\ref{fig:citations}.

As Figure~\ref{fig:citations} shows, the most ubiquous fields regarding citations are 
``Energy'', ``Chemical Engineering'' and ``Decision Sciences'' but this order is not related to the number 
of citations per field. Again, we can observe that the field with more citations is ``Medicine'' but it 
achieves only the 16-th place in terms of citation ubiquity, suggesting that the communities that develop 
this area are located in a few countries and produces many citations to other documents in this field. 
Something similar occurs in ``Biochemistry, Genetic and Molecular Biology'', field that has the 
lowest rank in terms of citation ubiquity but achieves the second place in terms of citations. 
Note that the closest fields of knowledge in this network are ``Chemistry'' and ``Materials Science''. 
Something similar occurs with ``Agricultural and Biological Sciences'' and ``Environmental Science''. 

\begin{figure}[h!]
\centering
\includegraphics[width=6.5cm]{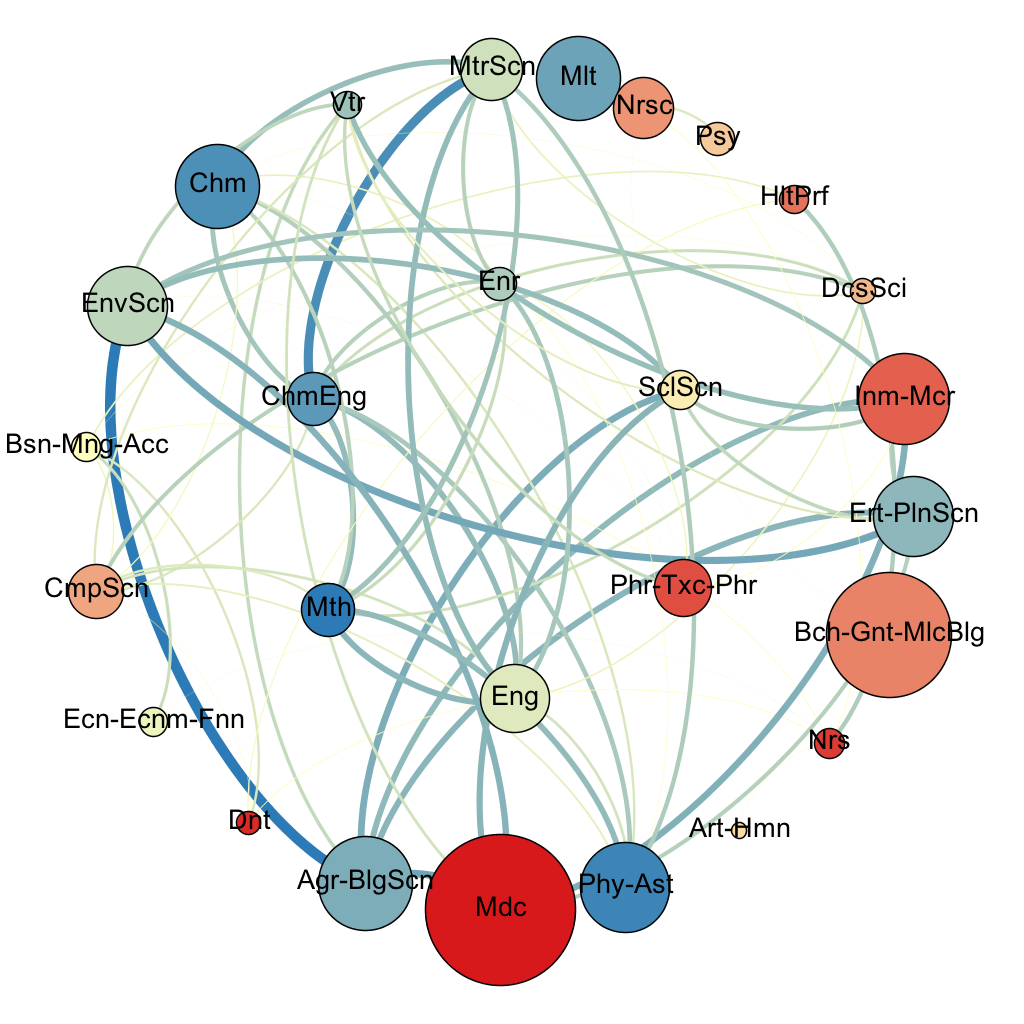}
\caption{A network visualization for RCA proximity based on the H-index.}
\label{fig:hindex}
\end{figure}

Finally, we visualize proximities between fields of knowledge according to the H-index. 
This network is shown in Figure~\ref{fig:hindex}.

As Figure~\ref{fig:hindex} shows, the node order (H-index ubiquity) is not related to node size. 
In fact, we can see that some of the most ubiquous fields in terms of H-index are also the ones with the 
lowest H-index values. We can see that the H-index is distributed across more fields than the previous 
indexes suggesting that H-index based-comparisons between fields of knowledge are more fair than the ones 
based on citations or number of documents. Finally, this network shows some semantic relationships. 
Relationships between ``Environmental Science'' and ``Agricultural and Biological Sciences'', 
and ``Chemistry'' and ``Materials Science'' arise as the most relevant of this network.

\section{Conclusions}

We present a RCA-based analysis for the world scientific production in the period that 
ranges 1995-2011 using the SCImago country and journal platform. The use of a RCA-based methodology 
favors the unleashing of comparative advantages between countries of different sizes and complexities. 
A number of visualizations and experimental results shows that the proposal is feasible and can be 
extended to conduct a global characterization at a high level of detail, for example, explaining the scientific 
complexity of each country. 

Currently we are conducting an analysis at country level, characterizing the complexity of each country 
in terms of its scientific production. Some preliminar results of this effort can be explored in our 
site~\footnote{\url{http://octopus.inf.utfsm.cl/~mguevara/KPS}}. In addition, we will explore the relation 
between the RCA-model and the term vector space model used in information retrieval, to include 
methodological improvements to this framework. 

\section{Acknowledgment}
Marcelo Mendoza was supported by project FONDECYT grant 11121435.

\bibliographystyle{abbrv}
\bibliography{csta}

\end{document}